\documentclass[10pt,conference]{IEEEtran}

\IEEEoverridecommandlockouts

\ifCLASSINFOpdf

\else
  
\fi

\usepackage{graphicx}
\usepackage{booktabs}
\usepackage{pifont}
\usepackage{url}
\usepackage{enumitem}

\newlist{customitemize}{itemize}{1}
\setlist[customitemize,1]{
  leftmargin=*, 
  label={\raisebox{-0.2ex}{\scalebox{1.4}{\textbullet}}}, 
  align=left 
}

\usepackage{algorithm}
\usepackage{algpseudocode}
\usepackage{amsmath}

\usepackage{hyperref}
\hypersetup{
	linkcolor=red,
	urlcolor=blue,
	citecolor=green,
}

\usepackage{bbding}
\usepackage{multirow}

\usepackage{tabularx}  

\usepackage{tcolorbox}
\newtcolorbox{mybox}[2][]{
  colback=gray!10,            
  colframe=black,             
  fonttitle=\bfseries\large,        
  coltitle=white,             
  colbacktitle=black,         
  title=#2,                   
  #1                          
}

\begin{document}

\title{PYPILINE: Malicious PyPI Package Detection via Suspicious API Knowledge and Agent Workflow}

\author{
\IEEEauthorblockN{
Siyuan Pang\IEEEauthorrefmark{2}\IEEEauthorrefmark{3},
Yepeng Yao\IEEEauthorrefmark{2}\IEEEauthorrefmark{3}\IEEEauthorrefmark{1}\thanks{\IEEEauthorrefmark{1} Corresponding author: Yepeng Yao},
Zhengwei Jiang\IEEEauthorrefmark{2}\IEEEauthorrefmark{3}\IEEEauthorrefmark{1}\thanks{\IEEEauthorrefmark{1} Corresponding author: Zhengwei Jiang},
Zijing Fan\IEEEauthorrefmark{2},
Haozhe Li\IEEEauthorrefmark{2}\IEEEauthorrefmark{3},
Baoxu Liu\IEEEauthorrefmark{2}\IEEEauthorrefmark{3}
} 

\IEEEauthorblockA{\IEEEauthorrefmark{2}\textit{Institute of Information Engineering, Chinese Academy of Sciences, Beijing, China}} 
\IEEEauthorblockA{\IEEEauthorrefmark{3}\textit{School of Cyber Security, University of Chinese Academy of Sciences, Beijing, China}}

\IEEEauthorblockA{\normalsize{\{pangsiyuan, yaoyepeng, jiangzhengwei,fanzijing, lihaozhe, liubaoxu\}@iie.ac.cn}}
}

\maketitle

\begin{abstract}
Detecting malicious PyPI packages is crucial for maintaining the security of the open source software supply chain. Traditional static rule detection methods require continuous maintenance by experienced security personnel, resulting in high labor costs. Dynamic analysis methods require actual execution of the target package code, posing a risk of malicious code proliferation, and incurring significant runtime overhead and low detection efficiency. Machine learning and LLM methods iterate the detection kernel but cannot invoke multiple tools, resulting in insufficient automation.To address these issues, we propose a novel detection method called PYPILINE, which combines suspicious API knowledge and agent workflow. PYPILINE first performs static analysis on known malicious packages, extracting abstract syntax trees and generating API call graphs. From these graphs, a structured suspicious API knowledge base is extracted and constructed. In the agent workflow, PYPILINE uses RAG technology to invoke this knowledge base to enhance analytical capabilities, performing in-depth semantic analysis of the packages, outputting structured evaluation reports, and automatically sending the reports to a mail server.Experimental results show that PYPILINE achieves precision of 96.7\%, recall of 99.6\%, and F1 score of 98.1\%. F1 score is improved by 5.7 to 21.6 percentage points compared to baseline tools. When 30 threads execute concurrently, detecting a single package takes an average of only 0.6 seconds.Furthermore, we conducted a large scale empirical study of malware packages, systematically revealing common attack strategies and the most frequently abused APIs. PYPILINE provides an intelligent, efficient, and automated package detection solution, enhancing the security of the open source software ecosystem.
\end{abstract}


\IEEEpeerreviewmaketitle

\section{Introduction} \label{sec:introduction}
Open source software (OSS) has become the cornerstone of modern software development. Synopsys 2024 Open Source Security and Risk Analysis report indicates that a staggering 96\% of codebases contain open source components \cite{introduction1}. However, the widespread adoption of OSS has also made its supply chain a high value target for attackers. Software package repositories \cite{huawei-pypi} \cite{tencent-pypi} \cite{alibaba-pypi}, as the central hubs of the supply chain \cite{pypi-index} \cite{pypi-simple}, are facing a severe challenge with the proliferation of malicious packages \cite{Neupane2023} \cite{Liu2022} \cite{Gu2023a} \cite{Gu2023}. According to a report by Sonatype, over 245,000 malicious packages were discovered in 2023 alone, a number exceeding the total from all years since 2019 \cite{introduction2}. Attackers employ tactics such as typosquatting \cite{qianxin-pypimal1} \cite{qianxin-pypimal2}, dependency confusion \cite{qianxin-pypimal3}, or hijacking legitimate accounts to embed malicious code into software packages \cite{qianxin-nugetmal1}. This code can be triggered during installation or use \cite{qianxin-pypimal4}, leading to severe consequences like data breaches and system compromise (e.g., Check Point CloudGuard found over 500 malicious packages via typosquatting attacks on PyPI) \cite{introduction3}.

Existing detection methods still have several shortcomings and cannot fully meet the high-efficiency detection needs of malicious packages in the PyPI ecosystem. First, traditional static rule based detection methods \cite{guarddog} \cite{OSSGadget} heavily rely on manually predefined heuristic rules, requiring continuous maintenance and updates by experienced security personnel, resulting in high labor costs. Furthermore, the rule iteration speed cannot keep up with rapidly evolving attack methods, significantly limiting adaptability and scalability. Second, dynamic analysis methods \cite{Duan2020maloss} require actual execution of the target package code to capture runtime malicious behavior, posing security risks of malicious code escape and propagation. They also have high resource consumption per package and low overall efficiency, making them unsuitable for rapid screening of massive package libraries. Third, while detection methods based on machine learning \cite{sejfia2022practical} \cite{chen2016xgboost} \cite{ke2017lightgbm} and large language models \cite{gao2025malguard} have iterated on the detection kernel and improved discriminative capabilities, most focus only on single-step classification tasks and cannot seamlessly integrate various external tools such as knowledge base retrieval and result push. The degree of automation throughout the process is insufficient, making it difficult to support a complete security operation loop.

To overcome these limitations, we propose a novel detection method: PYPILINE. The core innovation of PYPILINE lies in a tool-enabled LLM agent workflow. The designed agent can manipulate external tools, including retrieving refined suspicious API features from the vector database and delivering final detection reports to the mail server, enabling efficient, interpretable, and convenient malicious PyPI package detection. PYPILINE first performs static graph analysis on known malicious packages to automatically extract and build a structured suspicious API knowledge base. This systematically captures malicious behavioral patterns rather than relying on manual enumeration. During the online detection phase, the method leverages this knowledge base as key context to guide a Large Language Model based agent to conduct in depth semantic analysis and reasoning on the target package, ultimately outputting a structured, interpretable maliciousness assessment report. This knowledge enhanced AI workflow paradigm not only combines the efficiency of static analysis for pattern recognition with the advantages of automated knowledge extraction but also fully leverages the LLM's capabilities in complex code context understanding, logical reasoning, and natural language explanation. This achieves an excellent balance between detection performance and result interpretability.

We conducted a comprehensive evaluation of PYPILINE on a dataset containing 9,408 malicious packages and 14,005 benign packages. Experimental results show that PYPILINE significantly outperforms existing state of the art tools in terms of precision 96.7\%, recall 99.6\%, and F1-score 98.1\%. Furthermore, our empirical study on nearly ten thousand malicious packages systematically reveals the prevalence of mainstream attack strategies, such as install-time execution, code obfuscation, and downloading remote executables. It also provides statistics on the most frequently abused APIs, offering data driven insights into the threat landscape within the PyPI ecosystem. PYPILINE provides an intelligent, efficient, and automated detection solution for open source software supply chain security.

\textbf{Contributions.} The contributions of our paper are as follows:

$\bullet$ We propose \textbf{PYPILINE}, an intelligent AI agent workflow for malicious PyPI package detection based on suspicious API knowledge. It supports autonomous invocation of external tools including vector databases and mail servers, enabling an efficient and interpretable automated detection pipeline.

$\bullet$ We collected a dataset containing 5000 PYPI packages and used previously recognized datasets for testing. 

$\bullet$ Experiments demonstrate that our PYPILINE achieves the best results, outperforming previous tools.

$\bullet$ We thoroughly analyzed 23,413 known PYPI packages in the dataset, focusing particularly on suspicious APIs, malicious behavior classification, and interpretability analysis.

\section{Motivation} \label{sec:motivation}
\subsection{Background} 
With the widespread adoption of open source software (OSS) in software development, package repositories such as NPM and PyPI have become core infrastructure in the modern software supply chain. These repositories host millions of software packages, greatly improving development efficiency, but they have also become a breeding ground for software supply chain attacks. By planting malicious packages in these repositories, attackers can compromise downstream users and projects on a large scale at minimal cost.

The propagation of malicious packages primarily relies on several attack vectors, including: publishing entirely new malicious packages, conducting "typosquatting" attacks by imitating legitimate package names, and hijacking legitimate developer accounts or projects to tamper with existing packages. Regardless of the method, the ultimate goal of attackers is to have malicious code downloaded and executed by users. Depending on the stage of execution, malicious behaviors can be categorized into three types:

\textbf{Install-time attacks.} Exploiting the automatic execution of installation scripts by package managers (e.g., npm install, pip install). In NPM, attackers can embed malicious code in lifecycle hooks such as preinstallor postinstallin package.json; in PyPI, they can tamper with setup.pyscripts or insert malicious commands in cmdclass, causing the malicious code to execute immediately during installation.

\textbf{Import-time attacks.} When users import a module via require(Node.js) or import(Python), the entry file of the package (e.g., the file specified by the mainfield) is executed. Attackers can embed malicious initialization code in this file.

\textbf{Runtime attacks.} Malicious code is hidden within the functional functions of the package and only triggers under specific conditions or when users call particular functions, making it more stealthy.

\subsection{Case Study}

\begin{figure}
    \centering
    \includegraphics[width=\linewidth]{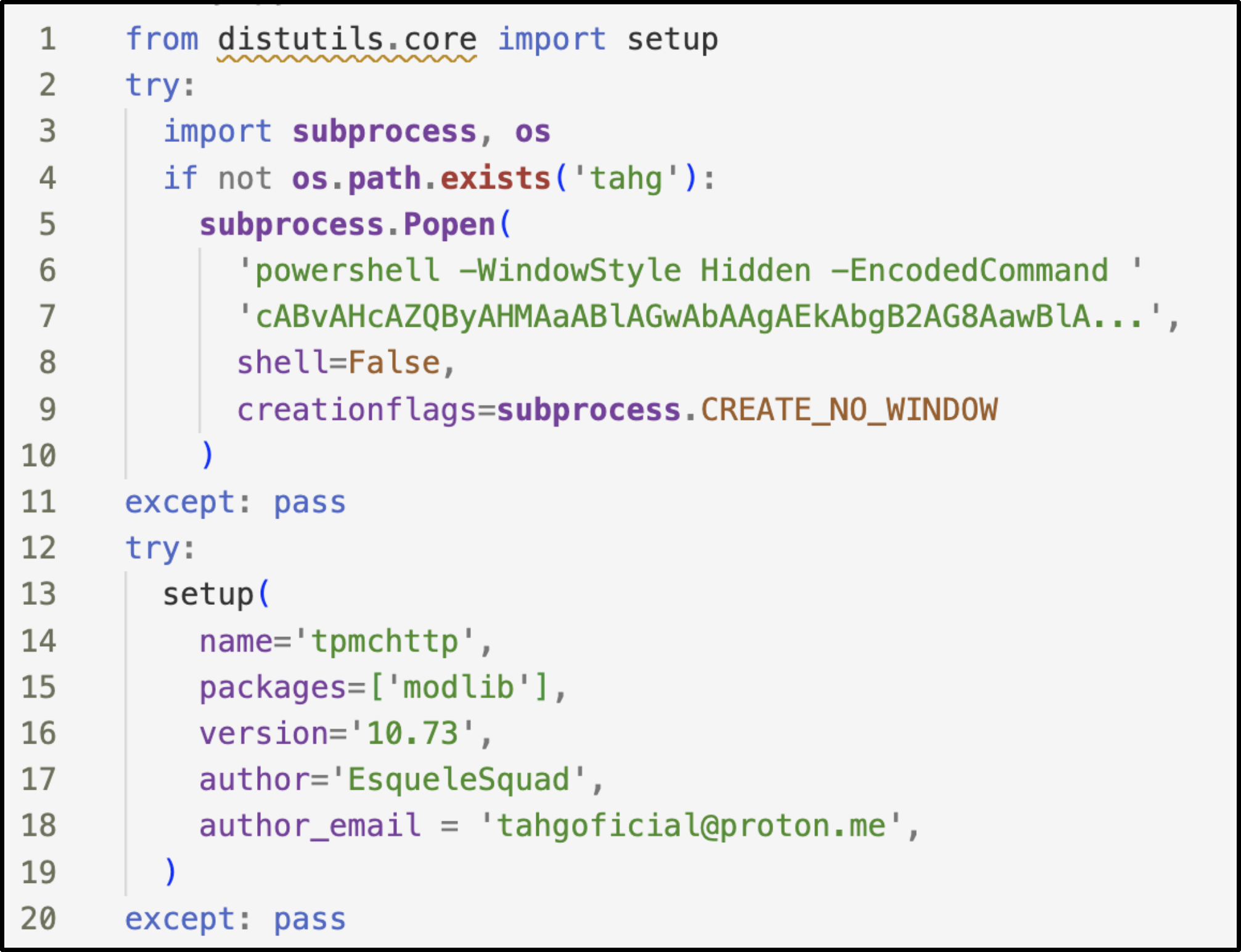}
    \caption{Case Study \#1}
    \label{case_study_1}
\end{figure}

\textbf{Case Study \#1:} Figure \ref{case_study_1} is a typical case of a software supply chain attack. The attacker embedded a malicious payload into the setup.py file, the core installation script of a Python package named tpmchttp. When a developer uses standard tools like pip install to install this package, setup.py is automatically executed as part of the installation process. Before executing the normal package metadata registration and installation logic (i.e., calling the setup() function), the script includes a preceding segment of malicious detection and execution code. This code first checks whether a file or directory named tahg exists in the current working directory as an attack trigger. If this marker is absent, the attack chain is initiated.

The core of the attack involves creating a hidden PowerShell process via Python's subprocess module. This process executes a Base64-encoded instruction. The decoded operation consists of: first, using the Invoke-WebRequest command to download a malicious executable named Esquele.exe from an address masquerading as a legitimate file-sharing service; subsequently, to enhance stealth, renaming it to WindowsCache.exe, which exhibits characteristics of a system file, and saving it to the user's home directory; finally, immediately executing this file using the Invoke-Expression command. The entire PowerShell process runs completely in the background without any user interface prompts, achieved through the -WindowStyle Hidden parameter and the CREATE\_NO\_WINDOW flag.

The intent of this technique is to exploit developers' trust in public software repositories by embedding a backdoor during critical stages of software construction or deployment. Upon successful installation, a remote malicious program (which could be ransomware, a remote access trojan, a cryptocurrency miner, or an information stealer) is downloaded and activated on the victim's system. This enables complete control of the target system or data theft, constituting a complete supply chain attack path from upstream software sources to end-users.

\begin{figure}
    \centering
    \includegraphics[width=\linewidth]{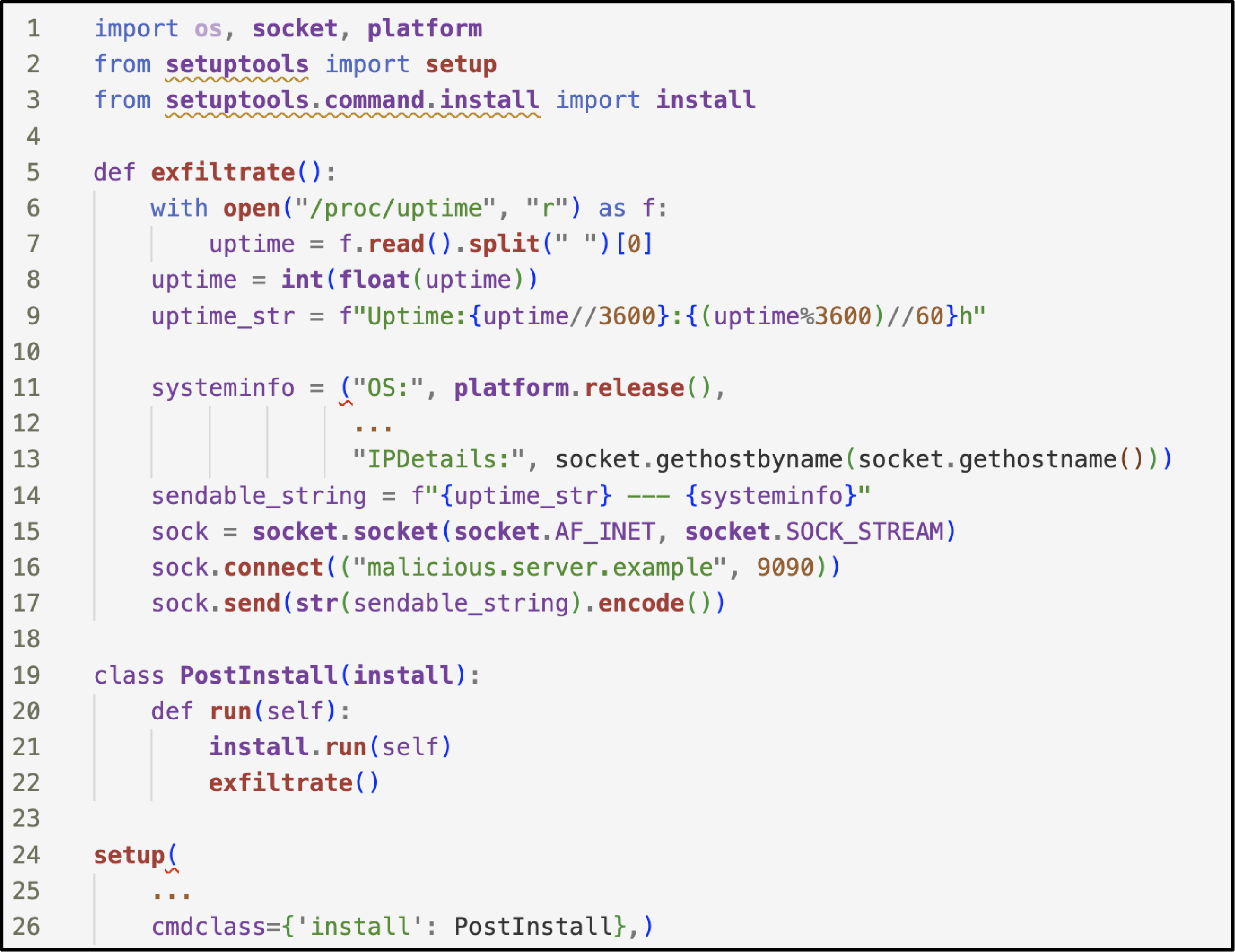}
    \caption{Case Study \#2}
    \label{case_study_2}
\end{figure}

\textbf{Case Study \#2:} This malicious code in Figure \ref{case_study_2} first reads the /proc/uptime file (Linux-specific) to obtain the system uptime, thereby determining whether the target is a long-running server worth targeting. Simultaneously, it uses the platform module to comprehensively collect operating system fingerprints, including kernel version (platform.release()), system name, detailed version, and full platform string. This precise information can be used for subsequent vulnerability matching. The code also retrieves the host's network address (typically the internal IP) via socket.gethostbyname(socket.gethostname()), helping the attacker locate the target within the network. After completing information gathering, its core action is to establish a raw TCP Socket connection, actively outbound to the attacker-controlled server and send the formatted system information package.

The entire malicious execution is cleverly embedded within the software installation process: the attacker defines a PostInstallCommand class, which inherits from the standard setuptools installation command and overrides the run method. This ensures that the information exfiltration function is automatically and silently invoked after the normal package installation concludes, with no abnormal prompts appearing in the user's terminal.

At its core, this malicious code is a lightweight information-stealing Trojan targeting Linux systems. While it lacks self-propagation or persistence capabilities, its functionality is highly focused: during the software installation moment, it silently conducts a rapid reconnaissance of the target host environment and exfiltrates critical metadata such as system fingerprints and network location via TCP to the attacker's server, providing intelligence support for potential follow-up attacks.

\begin{figure*}
    \centering
    \includegraphics[width=0.9\linewidth]{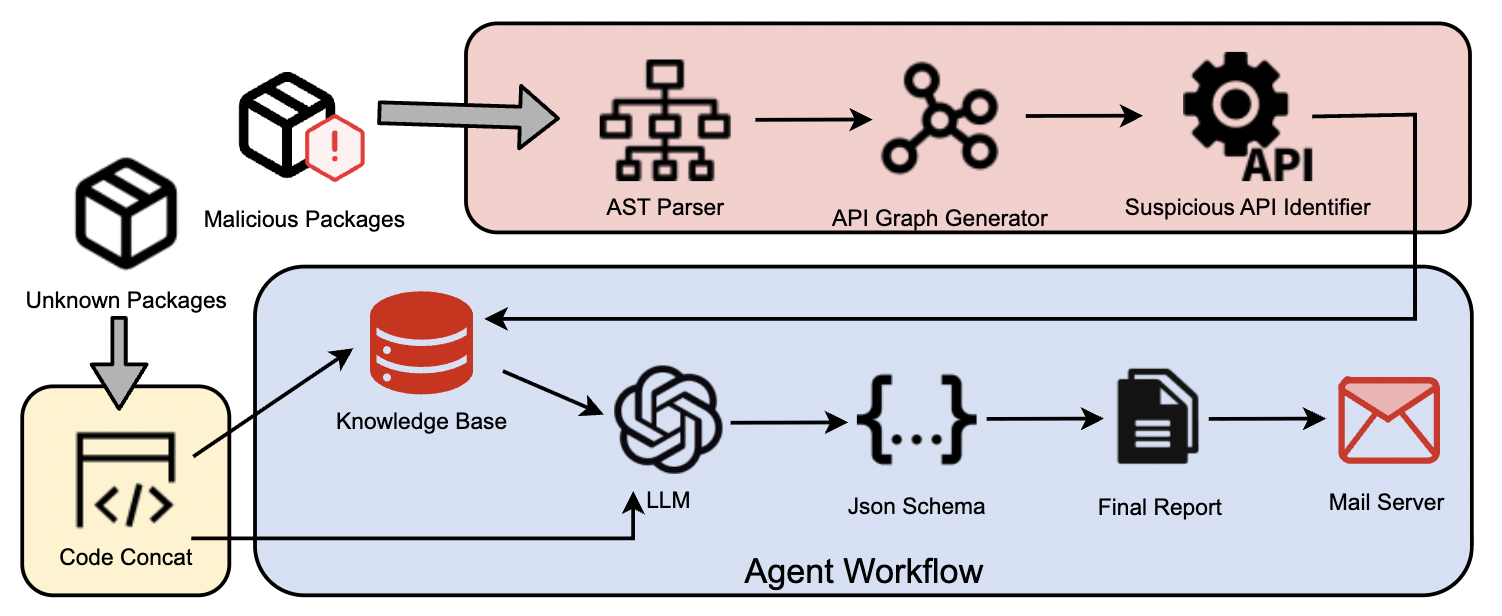}
    \caption{Overview of PYPILINE System}
    \label{system}
\end{figure*}

\section{METHODOLOGY} \label{sec:methodology}

To address the increasingly sophisticated and covert malware threats within the PyPI ecosystem, we propose \textbf{PYPILINE}. The core innovation of this method lies in constructing a structured knowledge base from sensitive API features extracted via graph analysis and deeply integrating it into an intelligent agent AI workflow. This enables efficient and explainable detection of maliciousness in unknown PyPI packages. Figure  \ref{system} illustrates the overall architecture of PYPILINE, whose workflow consists of two main phases: an offline knowledge base construction phase and an online detection phase. In the offline phase, a suspicious API knowledge base is automatically extracted from known malicious packages. In the online phase, this knowledge base is leveraged to enhance the reasoning capability of a large language model, which then analyzes and evaluates unknown packages through an intelligent workflow. Specifically, the intelligent agent is empowered to manipulate external tools: retrieving latent suspicious API knowledge from the built vector database and distributing the generated detection report to the dedicated mail server.

\subsection{Overview}
PYPILINE first performs static analysis on a dataset of known malicious PyPI packages, extracting abstract syntax trees and generating API call graphs. By calculating the degree centrality of nodes in these graphs, we quantify the structural importance of each API within a malicious context. Combined with insights from prior research, we filter and select the most representative set of suspicious APIs to construct a queryable knowledge base. During the detection phase, all Python source code files from the package under test are concatenated into a single document, which is then input—along with the suspicious API knowledge base—into a carefully designed large language model agent workflow. Guided by this knowledge base, the agent conducts in-depth semantic analysis of the code and ultimately outputs a structured maliciousness assessment along with corresponding explanations.

\subsection{AST Parser}
We use static analysis tools to parse each package in the offline malicious package dataset and construct its Abstract Syntax Tree. The AST accurately reflects the syntactic structure of the code. Based on the AST, we traverse all function call nodes to construct a package-level API call graph G = (V, E). Here, the node set V represents all APIs present in the code including standard libraries, third-party libraries, and user-defined functions, and the edge set E denotes the call relationships between APIs. If API A directly calls API B in the code, a directed edge exists from node A to node B.

\subsection{API Graph Generator}
This component aims to transform a flat list of API calls into a structured relational model, capturing the interactions and dependencies among APIs within the execution flow of malicious code. Its input is the API call list output by the previous component, and its output is a directed graph.

For a given API call sequence [api1, api2, api3, …], the graph generator constructs a directed graph G = (V, E) by executing the following steps: First, each unique API call name is added as a node in the graph. Next, edges are constructed based on the calling order in the code and potential dependency relationships. The specific strategy involves analyzing the context between API calls: if the execution result or behavior of one API caller directly or indirectly leads to the execution of another API callee, a directed edge is established from the caller to the callee. This process considers not only explicit function return value transfers but also implicit dependencies arising from side effects such as file operations, network activity, or process interactions. By doing so, the generated graph transcends a simple sequential list, reflecting critical functional chains and control flows in malicious code—for example, a complete attack chain from "network download" to "file write," and subsequently to "subprocess.Popen execution."

\subsection{Suspicious API Knowledge Generation}
The core objective of the knowledge base generator is to quantify and filter the most behaviorally indicative APIs from a large collection of malicious package API graphs. Its input consists of the API graph sets generated for all known malicious packages, and its output is a structured, maliciousness-correlation-ranked repository of suspicious APIs.

This component begins by performing centrality analysis on each API graph G in the collection. Centrality measures the importance of nodes within a graph. The system adopts degree centrality, calculated as the sum of in-degree and out-degree for each API node (e.g., os.system). In the context of malicious code, a higher degree centrality for an API indicates either a higher frequency of invocation within malicious functional modules or stronger associations with other malicious APIs, suggesting a greater likelihood of serving as a "hub" for malicious behavior. By statistically averaging the degree centrality across all malicious package graphs, a global structural importance score is derived for each API within the malicious ecosystem.

Subsequently, leveraging existing domain knowledge, the candidate API set ranked by degree centrality undergoes manual review and refinement. APIs that not only exhibit high centrality scores but have also been widely studied and confirmed to be strongly associated with malicious behavior are selected, forming the final repository of 300 malicious APIs. This knowledge base is stored in a structured format (API name, malicious behavior description) to provide precise, interpretable intelligence guidance for the subsequent online detection phase agent workflow.

\subsection{Code Concat}
To improve the processing efficiency of the large language model and provide complete package-level context, we first concatenate all .py source code files from the PyPI package under inspection into a single unified plain-text document, following the order of file paths. This process retains the original filenames as section headers, ensuring traceability of the code structure. This single document then serves as the foundational input for agent analysis.

\subsection{Agent Workflow Package Detection}
To enhance the automation and intelligence of detecting malicious code in Python third-party packages, PYPILINE designs an agent workflow driven by prompts. This workflow revolves around a meticulously designed prompt intended to guide large language models in conducting thorough security analysis of provided code.

First, the prompt explicitly defines the professional role of the large language model as a cybersecurity expert, thereby focusing its reasoning within the context of security analysis. Second, through contextual injection, it incorporates pre-constructed snippets of a malicious API knowledge base directly into the prompt as critical guidance. For example, the prompt instructs the model to pay special attention to whether the code includes known APIs associated with malicious activities, such as os.system, subprocess.Popen, eval, and opening files like etc/passwd. At the same time, the prompt clearly outlines the analysis task: analyzing all code within the Python Package to determine if it contains malicious behavior. It also provides the model with a specific analytical framework, covering aspects such as inspecting common malware patterns, assessing potential runtime hazards, maintaining analytical objectivity, and offering explanations.

To ensure that the analysis results can be efficiently parsed and integrated by automated systems, the prompt mandates that the large language model output detection results in a predefined and structured JSON format. This format strictly includes key fields like is\_malicious as a boolean judgment, confidence indicating the confidence level, mal\_types categorizing malicious types such as data theft, code obfuscation, or installation-time execution, api\_calls listing detected malicious APIs, and function providing a concise one-sentence summary of the malicious behavior. After careful consideration and weighing, as shown in Table \ref{malicious_types_classification}, this paper categorizes malicious code behaviors into eight types.

\begin{table}[]
\small
\centering
\caption{Classification of Various Malicious Types}
\label{malicious_types_classification}
\begin{tabularx}{\linewidth}{@{}c|c|X@{}}
\toprule
\textbf{ID} & \textbf{Malicious Category} & \textbf{Description} \\ 
\midrule
M1     & code-execution & Identify when an OS command is executed in the setup.py file.              \\
\midrule
M2     & obfuscation         & Identify when a package uses a common obfuscation method often used by malware.         \\
\midrule
M3     & exec-base64          & Identify when a package dynamically executes base64-encoded code.              \\
\midrule
M4     & download-executable         & Identify when a package downloads and makes executable a remote binary.           \\  
\midrule
M5     & dll-hijacking        & Identifies when a malicious package manipulates a trusted application into loading a malicious DLL.      \\
\midrule
M6     & shady-links          & Identify when a package contains an URL to a domain with a suspicious extension.      \\
\midrule
M7     & stealing-sensitive-data  & Identify when a package reads and exfiltrates sensitive data from the local system or the clipboard.   \\
\midrule
M8     & cmd-overwrite  & Identify when the 'install' command is overwritten in setup.py, indicating a piece of code automatically running when the package is installed.      \\
\bottomrule
\end{tabularx}%
\end{table}

During execution, PYPILINE submits the code documents for analysis together with the knowledge base-integrated prompt to the large language model, specifically the GPT5nano model in this workflow. By leveraging its deep semantic understanding of code and integrating the threat intelligence embedded in the prompt, the model performs comprehensive reasoning. Ultimately, it produces a complete structured JSON detection report. This report not only delivers a binary assessment of maliciousness but also details the confidence level, categorizes identified malicious types, lists specific malicious APIs triggered, and includes a natural language summary of their functionality. This method, which combines automated feature extraction with knowledge-enhanced AI reasoning, achieves high detection rates while providing semantic comprehension and rich interpretable outputs that traditional static analysis tools often lack. To realize full-process automation, the agent further invokes pre-defined functional tools: it retrieves correlated high-risk API features from the vector database to assist reasoning, and eventually submits the standardized JSON detection report to the mail server for subsequent security auditing and alerting.

\section{Evaluation} \label{sec:evaluation}
To evaluate PYPILINE, we investigate the following three research questions:

\begin{customitemize}[leftmargin=2em]
    \item \textbf{RQ1:} How effective is our tool PYPILINE in detection on the dataset?
\end{customitemize}

\begin{customitemize}[leftmargin=2em]
    \item \textbf{RQ2:} How do reducing the knowledge base and changing the LLM affect PYPILINE?
\end{customitemize}

\begin{customitemize}[leftmargin=2em]
    \item \textbf{RQ3:} How efficient is our tool in detection?
\end{customitemize}

\begin{customitemize}[leftmargin=2em]
    \item \textbf{RQ4:} Empirical Study of Malicious Behaviors in PYPI Packages.
\end{customitemize}

\subsection{Experimental Dataset}

\begin{table}[]
\large
\centering
\caption{Statistics of the constructed dataset}
\label{dataset}
\begin{tabular}{@{}l|cc@{}}
\toprule
\multicolumn{1}{l|}{\textbf{Dataset}} & \textbf{\#Malicious} & \textbf{\#Benign}\\
\midrule
Guo et al.\cite{guo2023empirical}         & 9,408  &-     \\
Gao et al.\cite{gao2025malguard}          & -      &9,005  \\
Ourwork                                   & -      &5,000  \\
\midrule
Total                                     & 9,408  &14,005 \\

\bottomrule
\end{tabular}%
\end{table}

\begin{table*}[htbp]
\normalsize
\centering
\caption{Comparison of PYPI Package Tools}
\label{tab:tool_comparison}
\begin{tabular}{@{}l|ccccccc@{}}
\toprule
\textbf{Tool} & \textbf{Malicious Types} & \textbf{Technique Used} & \textbf{Usability} & \textbf{Key Line Context} & \textbf{Explainability} & \textbf{Tool Call}\\
\midrule
\texttt{Guarddog} \cite{guarddog} & multi & Static (Rules) 
& \begin{tikzpicture}\fill[black] (0,0) circle (4pt);\end{tikzpicture} 
& \begin{tikzpicture}\draw (0,0) circle (4pt);\end{tikzpicture} 
& \begin{tikzpicture}\draw (0,0) circle (4pt);\end{tikzpicture} 
& \begin{tikzpicture}\draw (0,0) circle (4pt);\end{tikzpicture} 
\\

\texttt{bandit4mal} \cite{bandit} & multi & Static (Rules) 
& \begin{tikzpicture}\draw (0,0) circle (4pt);\fill[black] (0,0) --(-90:4pt) arc (-90:90:4pt) -- cycle;\end{tikzpicture} 
& \begin{tikzpicture}\draw (0,0) circle (4pt);\end{tikzpicture} 
& \begin{tikzpicture}\draw (0,0) circle (4pt);\end{tikzpicture} 
& \begin{tikzpicture}\draw (0,0) circle (4pt);\end{tikzpicture} 
\\

\texttt{OSSGadget} \cite{OSSGadget} & multi & Static (Rules) 
& \begin{tikzpicture}\draw (0,0) circle (4pt);\fill[black] (0,0) --(-90:4pt) arc (-90:90:4pt) -- cycle;\end{tikzpicture} 
& \begin{tikzpicture}\draw (0,0) circle (4pt);\end{tikzpicture} 
& \begin{tikzpicture}\draw (0,0) circle (4pt);\end{tikzpicture} 
& \begin{tikzpicture}\draw (0,0) circle (4pt);\end{tikzpicture} 
\\

\texttt{Malwukong} \cite{Li2023} & multi & Static (Rules)
& \begin{tikzpicture}\draw (0,0) circle (4pt);\end{tikzpicture} 
& \begin{tikzpicture}\fill[black] (0,0) circle (4pt);\end{tikzpicture} 
& \begin{tikzpicture}\draw (0,0) circle (4pt);\fill[black] (0,0) --(-90:4pt) arc (-90:90:4pt) -- cycle;\end{tikzpicture} 
& \begin{tikzpicture}\draw (0,0) circle (4pt);\end{tikzpicture} 
\\

\texttt{Maloss} \cite{Duan2020maloss} & binary & Static \& Dynamic 
& \begin{tikzpicture}\draw (0,0) circle (4pt);\end{tikzpicture} 
& \begin{tikzpicture}\draw (0,0) circle (4pt);\end{tikzpicture} 
& \begin{tikzpicture}\draw (0,0) circle (4pt);\end{tikzpicture} 

& \begin{tikzpicture}\draw (0,0) circle (4pt);\end{tikzpicture}
\\

\texttt{Ea4mp} \cite{sun20241+} & binary & Static (ML) 
& \begin{tikzpicture}\draw (0,0) circle (4pt);\end{tikzpicture} 
& \begin{tikzpicture}\draw (0,0) circle (4pt);\end{tikzpicture} 
& \begin{tikzpicture}\draw (0,0) circle (4pt);\fill[black] (0,0) --(-90:4pt) arc (-90:90:4pt) -- cycle;\end{tikzpicture} 
& \begin{tikzpicture}\draw (0,0) circle (4pt);\end{tikzpicture} 
\\

\texttt{Cerebro} \cite{zhang2025killing} & binary & Static (Bert) 
& \begin{tikzpicture}\draw (0,0) circle (4pt);\end{tikzpicture} 
& \begin{tikzpicture}\draw (0,0) circle (4pt);\end{tikzpicture} 
& \begin{tikzpicture}\draw (0,0) circle (4pt);\fill[black] (0,0) --(-90:4pt) arc (-90:90:4pt) -- cycle;\end{tikzpicture} 

& \begin{tikzpicture}\draw (0,0) circle (4pt);\end{tikzpicture} 
\\

\texttt{MalGuard} \cite{gao2025malguard} & binary & Static (ML) 
& \begin{tikzpicture}\draw (0,0) circle (4pt);\fill[black] (0,0) --(-90:4pt) arc (-90:90:4pt) -- cycle;\end{tikzpicture}  
& \begin{tikzpicture}\fill[black] (0,0) circle (4pt);\end{tikzpicture}  
& \begin{tikzpicture}\fill[black] (0,0) circle (4pt);\end{tikzpicture} 

& \begin{tikzpicture}\draw (0,0) circle (4pt);\end{tikzpicture} 
\\

\midrule

\texttt{PYPILINE} & multi & Static (Agent) 
& \begin{tikzpicture}\fill[black] (0,0) circle (4pt);\end{tikzpicture} 
& \begin{tikzpicture}\fill[black] (0,0) circle (4pt);\end{tikzpicture} 
& \begin{tikzpicture}\fill[black] (0,0) circle (4pt);\end{tikzpicture} 
& \begin{tikzpicture}\fill[black] (0,0) circle (4pt);\end{tikzpicture} \\
\bottomrule
\end{tabular}
\end{table*}

\textbf{Malicious Sample.} Since the detection capability and evaluation effectiveness of the methods benefit from large-scale and high-quality datasets, we construct our evaluation benchmark using a reliable, human-labeled dataset collected from the most reputable and authoritative paper by Guo et al.\cite{guo2023empirical} The detailed statistics of the dataset are shown in Table \ref{dataset}. After removing duplicate packages, the dataset includes a total of 9,408 malicious packages.

\textbf{Benign Sample.} We use the benign packages from the paper by Gao et al.\cite{gao2025malguard}, after removing duplicate packages which comprise 9,005 packages in total. Additionally, we randomly sample 5,000 popular packages from PyPI. A package is considered benign if it has been hosted on PyPI for more than 90 days and has more than 1,000 downloads.

\begin{table}[]
\normalsize
\centering
\caption{Results of Effectiveness Evaluation}
\label{tab:effectiveness}
\begin{tabular}{@{}l|ccc@{}}
\toprule
\textbf{Tool}  & \textbf{Precision (\%)} & \textbf{Recall (\%)} & \textbf{F1 (\%)} \\ \midrule
\texttt{OSSGadget} & 72.5          & 81.0           & 76.5   \\
\texttt{bandit4mal}  & 80.6          & 88.7           & 84.4   \\
\texttt{Guarddog}  & 86.9          & 78.6           & 82.5   \\
\texttt{MalGuard}  & 91.0          & 94.0           & 92.4   \\
\midrule
\texttt{PYPILINE}  & \textbf{96.7} & \textbf{99.6}  & \textbf{98.1}   \\ 
\bottomrule
\end{tabular}%
\end{table}

\subsection{RQ1: How effective is our tool PYPILINE in detection on the dataset?}

In the qualitative capability comparison shown in Table \ref{tab:tool_comparison}, we analyze the support of each tool across multiple dimensions. PYPILINE supports multiple types of malicious behavior detection, outperforming Maloss and Ea4mp, which only support binary classification. This indicates that PYPILINE covers a broader threat landscape. In terms of core technology, PYPILINE adopts a "static (agent)" approach, which is fundamentally different from the pure rule-based static analysis, static-dynamic hybrid, or machine learning-based methods employed by other tools. The agent workflow in PYPILINE enables more complex reasoning. Regarding the level of detail and comprehensibility of output, PYPILINE demonstrates significant advantages: it can pinpoint risky code and provide contextual code snippets for critical lines. More importantly, thanks to the integration of large language models, PYPILINE offers comprehensive support for explainability, generating human-readable detection rationales and analyses. In contrast, most comparative tools either cannot provide explanations or offer only limited ones. PYPILINE exhibits good usability; we have successfully implemented and deployed it for real-world operation, achieving excellent results. More importantly, PYPILINE is the only tool that uses an intelligent agent workflow, which can intelligently access databases and mail servers, further automating the malware detection process.

In the quantitative performance evaluation shown in Table \ref {tab:effectiveness}, we compared the detection performance of PYPILINE with four open-source detection tools (OSSGadget, bandit4mal, Guarddog, and MalGuard) on the same dataset. Experimental results show that PYPILINE achieves the best overall performance. Its precision reaches 96.7\%, significantly higher than all other comparison tools 72.5\%–91.0\%, indicating that PYPILINE generates very few false positives; its recall rate is as high as 99.6\%, also better than the other baseline methods 78.6\%–94.0\%, demonstrating extremely strong malicious package capture capabilities. MalGuard's overall performance is better than OSSGadget, bandit4mal, and Guarddog, with precision, recall, and F1 score of 91.0\%, 94.0\%, and 92.4\%, respectively, but its overall detection performance still lags significantly behind PYPILINE. PYPILINE achieved a final F1 score of 98.1\%, significantly outperforming second-place MalGuard. This result demonstrates that PYPILINE analysis method and agent workflow, based on knowledge of suspicious APIs, achieves an optimal balance between detection accuracy and reliability.

\vspace{5pt} 
\noindent 
\colorbox{gray!30}{
\parbox{\dimexpr\columnwidth-2\fboxsep\relax}{
\textbf{Summary:} The evaluation results strongly validate the effectiveness of our tool PYPILINE in detecting malware packages on PyPI. It not only provides more detailed risk identification and functional explanations but also intelligently invokes tools. PYPILINE also leads in key performance indicators, effectively reducing false positives and false negatives.
}}
\vspace{5pt} 

\subsection{RQ2: How do reducing the knowledge base and changing the LLM affect PYPILINE?}

\begin{table}[]
\normalsize
\centering
\caption{Evaluation of Different Chinese LLM in Agent Workflow}
\label{tab:different_LLM}
\begin{tabular}{@{}l|ccc@{}}
\toprule
\textbf{Tool}  & \textbf{Precision (\%)} & \textbf{Recall (\%)} & \textbf{F1 (\%)} \\ \midrule
\texttt{PYPILINE} & \textbf{96.7}          & \textbf{99.6}           & \textbf{98.1}   \\  \midrule

\texttt{PYPILINE(ds)} & 50.3          & 97.8           & 66.4   \\
\texttt{PYPILINE(km)}  & 54.9          & 93.3           & 69.1   \\
\texttt{PYPILINE(db)}  & 39.2          & 43.4           & 41.2   \\
\texttt{PYPILINE(glm)}  & 42.0          & 61.7           & 50.0   \\

\bottomrule
\end{tabular}%
\end{table}

\begin{table}[]
\small
\centering
\caption{Evaluation after removing knowledge base}
\label{tab:w/o_knowledge}
\begin{tabular}{@{}l|ccc@{}}
\toprule
\textbf{Tool}  & \textbf{Precision (\%)} & \textbf{Recall (\%)} & \textbf{F1 (\%)} \\ \midrule
\texttt{PYPILINE} & \textbf{96.7}          & \textbf{99.6}           & \textbf{98.1}   \\   \midrule

\texttt{(w/o) Knowledge} & 92.9          & 97.3      & 95.0  \\

\bottomrule
\end{tabular}%
\end{table}

RQ2 systematically evaluates the impact of model selection and removing knowledge base on the performance of the PYPILINE tool through two sets of controlled experiments.

As shown in Table \ref{tab:different_LLM}, focuses on the replacement testing of the agent's internal model. To assess PYPILINE's potential adaptability in the Chinese mainland context, we replaced its natively deployed gpt5nano model with four mainstream Chinese domestic LLMs and ran them on the same test set. The results, show a significant performance gap. Specifically, the deepseek-v3.2 model \cite{liu2024deepseek} demonstrated a very high recall (97.8\%), but its precision was only 50.3\%, leading to a high number of false positives and a final F1-score of 66.4\%. The kimi-k2 model \cite{team2025kimi} achieved a relatively better balance between precision (54.9\%) and recall (93.3\%), with an F1-score of 69.1\%, representing the best performance among the domestic models, but it still lags significantly behind the baseline. The performance metrics for doubao-2.0-mini \cite{doubao} and glm-4.7 models \cite{du2022glm} were generally lower; for example, both the precision and recall of doubao-2.0-mini were below 45\%. In stark contrast, the native PYPILINE configuration achieved near-perfect scores across all three metrics, with 96.7\% precision, 99.6\% recall, and an F1-score of 98.1\%. The significant advantage of international mainstream models may stem from their targeted training and optimization on massive, multi-turn code and security-related data, enabling them to accurately understand the contextual semantics of malicious API sequences. The Chinese domestic models tested, while potentially effective in general tasks, showed clear deficiencies in the specialized task of malicious package detection, which requires risk awareness. Their decision-making granularity, knowledge coverage of attack patterns, and robustness against adversarial samples were significantly lacking, making them prone to false positives or false negatives.

As shown in Table \ref{tab:w/o_knowledge}, compared to the original PYPILINE tool with a complete knowledge base, removing the malicious API knowledge base information resulted in several significant drops: Recall decreased from 99.6\% to 97.35\%; Precision decreased from 96.7\% to 92.85\%; and the overall F1 score dropped from 98.1\% to 95.05\%. The removal of the knowledge base significantly weakened PYPILINE's ability to characterize and capture various malicious behaviors, leading to an increase in both false negatives of malicious samples and false positives of benign samples.

\vspace{5pt} 
\noindent 
\colorbox{gray!30}{
\parbox{\dimexpr\columnwidth-2\fboxsep\relax}{
\textbf{Summary:} A high-performance LLM ensures the reliability of malicious package identification, providing a solid foundation for the agent workflow. The knowledge base provides crucial prior knowledge for malicious detection and is a vital support module ensuring PYPILINE's high accuracy and broad coverage detection capabilities. Future work could explore fine-tuning methods for more Chinese mainland models on the problem of malicious package detection to better adapt to different environmental requirements. 
}}
\vspace{5pt} 

\subsection{RQ3: How efficient is our tool in detection?}

\begin{table}[]
\normalsize
\centering
\caption{time taken to process a single batch by different Workers}
\label{tab:time_taken}
\begin{tabular}{@{}cccc@{}}
\toprule
\textbf{Batch} & \textbf{Max Workers} & \textbf{Total Time} &\textbf{Time per Package}\\ \midrule

\#1 & 5          & 3319s           & 18.0s   \\
\#2 & 10         & 2202s           & 22.0s   \\
\#3 & 20         & 1140s          & 22.8s   \\
\#4 & 30          & 603s           & 18.1s   \\  
\bottomrule

\end{tabular}%
\end{table}

As shown in Table \ref{tab:time_taken}, shifts the focus to system efficiency, investigating the impact of the hyperparameter "maximum worker threads" on the processing speed of the PYPILINE pipeline. We fixed a batch size to 1000 packages for detection and varied the concurrency by adjusting the max\_workers parameter of the ThreadPoolExecutor (5, 10, 20, 30). The total processing time decreased almost linearly and substantially as the number of worker threads increased: from 3319 seconds with 5 threads to 603 seconds with 30 threads, achieving a speed-up factor exceeding 5.5x. This intuitively demonstrates the good parallelizability of tasks within the PYPILINE workflow; increasing concurrency effectively utilizes multi-core computational resources, significantly improving the throughput of batch operations. However, another key metric, the "average processing time per package," exhibited a different trend, fluctuating slightly between 18.0 and 22.8 seconds without showing a systematic decrease as the thread count increased. This reveals the nature of the performance bottleneck: the analysis time for a single software package is primarily consumed by core computational steps such as LLM inference and API knowledge base matching. The duration of these steps is determined by the model and algorithmic requirements and is relatively fixed. The concurrency optimization addresses task queuing and scheduling overheads rather than shortening the inherent "thinking" time of each task. This test provides important tuning guidance for practical deployment: users can flexibly configure the max\_workers parameter based on their hardware resources and task urgency to maximize efficiency within an acceptable time frame.

\vspace{5pt} 
\noindent 
\colorbox{gray!30}{
\parbox{\dimexpr\columnwidth-2\fboxsep\relax}{
\textbf{Summary:} Increasing the number of max\_workers can optimize resource utilization through parallelization, thereby shortening the overall analysis time. This allows PYPILINE to provide a practical solution for malware detection, a security application scenario, with high efficiency.
}}
\vspace{5pt} 

\subsection{RQ4: Empirical Study of Malicious Behaviors in PYPI Packages.}

To empirically study the behavioral patterns of malicious activities in the PyPI ecosystem, this research conducted a detailed analysis of a dataset comprising nearly 10,000 malicious PyPI packages, systematically quantifying the strategies commonly employed by attackers. The observed malicious behaviors in the code were categorized into eight core types, with their definitions and statistical counts shown in the accompanying Table \ref{mal_types}.

\begin{table}[]
\normalsize
\centering
\caption{Statistics of Malicious Behavior Types}
\label{mal_types}
\begin{tabular}{@{}c|l|c@{}}
\toprule
\multicolumn{1}{l|}{\textbf{ID}} &\textbf{Malicious Behaviour Types} & \textbf{Counts} \\
\midrule
M1 & code-execution-during-installation & 8352   \\
M2 & obfuscation                        & 6510   \\
M4 & download-executable                & 6742   \\
M3 & exec-base64                        & 5639   \\
M6 & shady-links                        & 3486   \\
M7 & stealing-sensitive-data            & 821    \\
M8 & cmd-overwrite                      & 87     \\
M5 & dll-hijacking                      & 6      \\
\bottomrule
\end{tabular}%
\end{table}

Payload delivery and immediate execution are primary objectives, with code-execution-during-installation (8,352 instances) and download-executable (6,742 instances) being the two most prevalent malicious behaviors. This highlights the attackers' core strategy: leveraging the trusted and automated process of package installation as a "trigger" to initiate the attack chain. The former directly embeds malicious commands, while the latter implements remote delivery of "staged payloads." These two methods are often combined to maximize the success rate of attacks.

Evasion techniques to bypass detection have become standard practice. The widespread use of obfuscation (6,510 instances) and exec-base64 (5,639 instances), covering 65\% and 56\% of the samples respectively, demonstrates that employing technical methods to evade detection by security tools has become a "basic operation" for malicious packages. This directly confirms the evolution of attackers' strategies—while achieving malicious functionality, they must prioritize bypassing detection solutions based on signatures or static patterns.

The frequent occurrence of shady-links (3,486 instances) reveals commonalities in attack infrastructure. Attackers tend to avoid hardcoding the final payload within the package, instead dynamically retrieving instructions from suspicious domains. This provides attackers with significant flexibility to remotely update attack targets, replace payloads, or initiate different attack stages after the package is released, forming a modular attack model of "dropper + remote C2."

Although exfiltrate-sensitive-data (821 instances) appears in lower absolute numbers compared to execution-based behaviors, its presence confirms that data theft is one of the core motivations for attacks. "Executing malicious code" is often the means, while "exfiltrating sensitive data such as system information and credentials" is a common and direct objective for monetization or reconnaissance. Cmd-overwrite (87 instances) and dll-hijacking (6 instances), despite their smaller sample sizes, represent more covert and targeted advanced threats. Cmd-overwrite involves deeper hijacking of the installation process, while dll-hijacking leverages operating system mechanisms for privilege persistence. The emergence of these techniques indicates the presence of a small number of more technically capable attackers in the PyPI ecosystem, whose behaviors are shifting from "broad distribution" to "precision infiltration."

\vspace{5pt} 
\noindent 
\colorbox{gray!30}{
\parbox{\dimexpr\columnwidth-2\fboxsep\relax}{
\textbf{Summary:} This empirical study quantitatively reveals the priorities and combination patterns of attack strategies in malicious PyPI packages. The core logic of attackers is to forcibly or automatically trigger commands during the installation phase in a highly covert manner, achieving the delivery and execution of remote control payloads, and ultimately accomplishing malicious objectives such as data exfiltration. This provides clear data support for building more targeted detection models in the future, suggesting that defenders should focus on installation-time behaviors, dynamic decoding techniques, and connection attempts to external suspicious domains.
}}
\vspace{5pt} 

\begin{table}[]
\normalsize
\centering
\caption{Suspicious API Calls Top15}
\label{mal_api_calls}
\begin{tabular}{@{}c|l|c@{}}
\toprule
\multicolumn{1}{c|}{\textbf{No.}} & \textbf{Suspicious API Calls} & \textbf{Counts} \\
\midrule
\#1 & os.path.exists                      & 5557   \\
\#2 & subprocess.Popen                    & 5280   \\
\#3 & subprocess.CREATE\_NO\_WINDOW         & 4351   \\
\#4 & open                                & 1173   \\
\#5 & exec                                & 1137   \\
\#6 & Invoke-WebRequest (Powershell)                   & 1047   \\
\#7 & Invoke-Expression (Powershell)                   & 1028   \\
\#8 & os.getenv                           & 1011   \\
\#9 & distutils.core.setup                & 977    \\
\#10 & powershell -EncodedCommand          & 934    \\
\#11 & os.makedirs                         & 899    \\
\#12 & os.walk                             & 888    \\
\#13 & pip.main                            & 861    \\
\#14 & ctypes.windll.shell32.IsUserAnAdmin & 789    \\
\#15 & os.path.expanduser                  & 783   \\
\bottomrule
\end{tabular}%
\end{table}

As shown in the Table \ref{mal_api_calls} Frequent occurrences of os.path.exists and os.getenv reveal the starting point of the attack chain. Before executing malicious operations, attackers prioritize environmental reconnaissance by checking specific paths, files, or environment variables to determine whether they are in a sandbox or if the target machine meets the attack conditions. This helps them decide whether to trigger subsequent malicious activities, thereby enhancing the targeting and evasion capabilities of the attack.

subprocess.Popen, as the absolute mainstream command execution interface, directly supports the most frequent behavior—"code-execution-during-installation." Its frequent combination with the subprocess.CREATE\_NO\_WINDOW flag quantitatively confirms attackers’ extreme pursuit of stealth—ensuring malicious processes run silently in the background without alerting end users. The frequent use of exec forms the technical foundation for "exec-base64" behavior. Attackers encode malicious code, dynamically decode it, and execute it, effectively evading detection schemes based on static string matching.

The collective appearance of PowerShell related calls Invoke-WebRequest, Invoke-Expression, and powershell -EncodedCommand each exceeding 930 occurrences, highlights that PowerShell has become the preferred and powerful toolchain for remote delivery (download-executable) and obfuscated execution in malicious packages on the Windows platform. The use of the -EncodedCommand parameter further demonstrates evasion techniques implemented at the parameter level.

The frequent appearance of os.walk and os.path.expanduser provides specific technical annotation for "exfiltrate-sensitive-data" behavior. Attackers systematically search for and exfiltrate credentials, keys, and personal data by traversing the user’s home directory and the entire disk. The abuse of distutils.core.setup indicates that attackers deeply exploit the standard installation entry point of PyPI packages. The check for ctypes.windll.shell32.IsUserAnAdmin shows that malicious packages possess permission awareness, preparing for deeper system-level attacks (such as persistence).

\vspace{5pt} 
\noindent 
\colorbox{gray!30}{
\parbox{\dimexpr\columnwidth-2\fboxsep\relax}{
\textbf{Summary:} The statistical analysis of malicious API calls concretizes macro behavioral classifications into observable and countable technical methods at the code implementation level. The data indicates that attackers’ technical implementations revolve around a core chain of silent reconnaissance, stealthy execution, abuse of legitimate tools (especially PowerShell), dynamic obfuscation, and targeted data collection.
}}
\vspace{5pt} 

\section{DISCUSSION} \label{sec:discussion}
\subsection{Challenges and Limitations}
Although our approach shows promising potential in experiments, it still faces a series of challenges and limitations in practical deployment, which are key directions for future improvements.

\textbf{Trade-offs in Model Selection.} Large language models are the core of the workflow, and their selection directly determines the system's performance and cost. We observed significant performance differences among various models in malware detection tasks. While the latest and most capable models (such as the GPT-5 series) generally achieve higher accuracy, their API call costs are significantly higher. In contrast, some open-source or smaller-scale models may be low-cost or even free, but their judgment accuracy and stability are insufficient, making them more prone to false negatives or false positives. Additionally, different models exhibit distinct "personality" biases; for example, some models may lean toward classifying ambiguous code as malicious, while others may exhibit excessive caution by frequently outputting "uncertain" conclusions. This non-determinism poses challenges for building stable and reliable industrial-grade systems.

\textbf{Reliability of Output Formatting.} To enable automated integration with downstream systems, we require large models to output structured reports in strict JSON format. However, despite providing detailed examples in the prompts, models can still generate malformed output, such as incorrect key names, mismatched value types, JSON syntax errors, or additional explanatory text appended before or after the JSON, especially for deep-thinking large models. This "formatting hallucination" often causes pure JSON parsers to fail. To address this, additional steps must be added to the agent workflow to clean and repair the output.

\textbf{Model Hallucination and Misjudgment.} The inherent hallucination issue in large language models is particularly hazardous in code security auditing. Specific manifestations include: Fabricated malicious behavior, a model may classify an inherently benign software package as malicious and "imagine" a seemingly plausible but non-existent attack chain to support its conclusion. Category overflowz, a model may output malicious code type labels outside the predefined classification, despite explicit restrictions in the output options. Overinterpretation of ambiguous code, certain legitimate code with high-privilege operations (such as network access or file writing) may be misjudged as malicious by the model based on statistical correlations. These issues lead to a certain rate of false positives.

\subsection{Future Work Outlook}
Promoting the Practical Deployment and Application of Tools. The ultimate goal of this research is to enhance the security level of the open-source software ecosystem. We call for and hope to collaborate with new software supply chain security startups or relevant departments of large security enterprises to transform such research tools into mature, robust commercial or community-driven products. An ideal application scenario would involve a professional team operating this tool to regularly scan mainstream open-source repositories such as PyPI and NPM, and release authoritative reports and alerts on malicious software packages, thereby forming public or commercial services. This would not only translate academic achievements into practical productivity but also, through large-scale and continuous application, collect more data and feedback to further drive the iteration and innovation of detection technologies.

\section{Related Work} \label{sec:related work}

\subsection{Malicious Package Detection.}
Malicious package detection is a core issue in software supply chain security, aiming to identify packages containing malicious code uploaded to package registries (e.g., npm, PyPI). Existing methods can be categorized into rule-based, machine learning based, and large language model based approaches \cite{yu2024maltracker} \cite{guo2026understanding}.

Rule-based methods rely on predefined rules or features (e.g., sensitive API calls, metadata anomalies) to flag suspicious packages. For instance, GUARDDOG \cite{guarddog} uses static analysis and heuristic rules to detect malicious behavior; OSSGadget \cite{OSSGadget} identifies potential backdoors by analyzing package metadata and code patterns. However, these methods depend on manual feature engineering, struggle to adapt to rapidly evolving attacks, and often suffer from high false positive rates.

Machine learning methods employ feature extraction and classification models \cite{ke2017lightgbm} \cite{freund1997decision} (e.g., Random Forest \cite{breiman2001random}, XGBoost \cite{chen2016xgboost}) for detection. For example, AMALFI \cite{sejfia2022practical} trains classifiers on features like API sequences and metadata extracted from packages; CEREBRO \cite{zhang2025killing}  fine-tunes BERT models based on code behavior sequences to capture semantic information. While these methods improve detection precision, they remain limited by the comprehensiveness of the feature set and are less effective against obfuscated code.

LLM-based approaches have gained significant attention recently, utilizing prompt engineering or fine-tuned LLMs (e.g., GPT, BERT) to automatically analyze code behavior. SocketAI \cite{zahan2025leveraging} employs an iterative prompting technique, enabling the LLM to perform multi-round reviews of potentially malicious code; SPIDERSCAN \cite{huang2024spiderscan} leverages LLMs to identify sensitive APIs and constructs a graph-based behavior model for pattern matching. LLMs demonstrate a clear advantage in understanding code context but face challenges such as hallucination and high computational costs.

Some research also combines static and dynamic analysis to enhance robustness. DONAPI \cite{huang2024donapi} detects obfuscated packages by reconstructing code dependencies and extracting dynamic behavior; MALGUARD \cite{gao2025malguard} integrates graph centrality analysis with the LIME algorithm to provide explainable outputs. These methods attempt to balance detection efficiency and accuracy, though dynamic analysis often introduces performance overhead.

\subsection{Package Registry Security.}

Package registry security focuses on vulnerabilities and attack vectors inherent to registry platforms (such as npm, PyPI, and Docker Hub), including supply chain attacks such as dependency obfuscation and typosquatting. Empirical studies show that the openness and automated processes of registries are easily exploited by malicious actors \cite{ladisa2023feasibility} \cite{wu2025exposing} \cite{miller2025understanding}. Different registries have varying security mechanisms, leading to attack migration risks. Language-independent detection methods need to be developed, but existing tools are mostly platform-specific and lack universality.

For npm and PyPI registries, common attacks include typosquatting (registering malicious packages with names similar to popular packages to trick users into downloading them), dependency obfuscation (uploading public packages with the same names as private packages to hijack dependency installation), and account hijacking (implanting malicious code through social engineering or credential leaks). For example, in the npm node-ipc \cite{node-ipc} incident, attackers used typosquatting to distribute malicious packages that overwrote disk files.

Containers face various quality issues, such as being outdated, containing vulnerabilities, leaking secret content, or being inherently malicious \cite{wong2023security}. Many studies have been conducted based on this \cite{rosa2023quality} \cite{henkel2021shipwright}. Dahlmanns et al. \cite{dahlmanns2023secrets} investigated the issue of including private keys or API secrets in container images. Zerouali et al. \cite{zerouali2019relation} discussed the relationship between outdated containers and the vulnerable packages they contain \cite{zerouali2019impact}. Tanaka et al. \cite{tanaka2023meta} explored the applicability of the Dockerfiles meta-maintenance approach. The container registry also faces security risks. Liu et al. \cite{liu2022exploring} conducted the first in-depth study on container registries' vulnerabilities to typosquatting attacks, proposing a lightweight extension to image management but not identifying any poisoned images. Liu et al. \cite{liu2020understanding} revealed and analyzed Docker Hub's security issues through the first large-scale analysis, although they only used static analysis to find malicious images and detect programs that had already been executed in containers.

\section{Conclusion} \label{sec:conclusion}
In this paper, we propose PYPILINE, a novel approach for malicious PyPI package detection that integrates a suspicious API knowledge with a agent workflow. The results demonstrate that PYPILINE significantly outperforms existing state of the art tools, achieving superior precision of 96.7\%, recall of 99.6\%, and F1-score of 98.1\%. Furthermore, the agent workflow exhibits good scalability; by adjusting the concurrency parameter, the system's throughput can be efficiently optimized, although the core analysis time per package remains constant.  We conducted an empirical study on malicious packages, quantitatively revealing the prevalence of attack strategies like installation time execution and obfuscation, as well as the most frequently abused APIs. PYPILINE represents a practical step towards deploying intelligent, efficient, and automated detection systems to enhance the security of the open source software supply chain.

\section*{Data Availability}
The code for the PYPILINE tool can be publicly accessed at 
\url{https://doi.org/10.5281/zenodo.19342665}


\newpage

\bibliographystyle{IEEEtran}
\bibliography{IEEEabrv,references}

\end{document}